\documentclass[12pt]{iopart}

\usepackage{iopams}
\usepackage{graphicx}
\newcommand{\bra}[1]{\left\langle{#1}\right\vert}
\newcommand{\ket}[1]{\left\vert{#1}\right\rangle}

\begin{document}

\paper{Cluster states from imperfect global entanglement}
\author{Michael C. Garrett$^1$ and David L. Feder$^2$}
\address{$^1$ ARC Centre of Excellence for Quantum-Atom Optics and School of Physical Sciences, University of Queensland, Brisbane, Australia 4072}
\address{$^2$ Institute for Quantum Information Science and Department of 
Physics and Astronomy, University of Calgary, Calgary, Canada T2N 1N4}
\eads{\mailto{mgarrett@physics.uq.edu.au}, \mailto{feder@phas.ucalgary.ca}}

\begin{abstract}
The cluster state, the highly entangled state that is the central
resource for one-way quantum computing, can be efficiently generated in a
variety of physical implementations via global nearest-neighbor interactions.
In practice, a systematic phase error is expected in the entangling process,
resulting in imperfect cluster states. We present a stochastic measurement
technique to generate large perfect cluster states and other graph states with
high probability from imperfect cluster states even when their initial
entanglement is weak.
\end{abstract}
\pacs{03.67.Lx, 03.67.Mn, 32.80.Pj}
\submitto{\NJP}
\tableofcontents
\title[Cluster states from imperfect global entanglement]{}
\maketitle

\section{Introduction}
One-way quantum computing (1WQC)~\cite{Raussendorf2001,Raussendorf2003a} boasts the advantage
over the standard quantum circuit approach of allowing all entanglement to be
prepared in a single initial step prior to any logical operations. This
initial resource, known as the ``cluster state,'' is a highly entangled
multipartite state~\cite{Briegel2001}. Universal quantum computation is then
achieved through single-qubit measurements alone, thereby eliminating the
troublesome requirement for dynamically controlled two-qubit operations.

The cluster state can be efficiently produced from a physical lattice of
qubits, each initialized in the state
$\ket{+}=\left(\ket{0}+\ket{1}\right)/\sqrt{2}$,
by applying controlled-$\sigma_z$ ($CZ$) operators between all nearest-neighbor qubit pairs, where $\sigma_x$, $\sigma_y$, and $\sigma_z$ are the standard Pauli matrices. In principle, this can be achieved by a combination of Ising interactions and
external fields, corresponding to a Hamiltonian of the form
\begin{equation} \label{eq:ham}
  H = g \sum_{i<j}{\left(\frac{\sigma_z^{(i)}-1}{2}\right)\left(\frac{\sigma_z^{(j)}-1}{2}\right)} \mathrm{,}
\end{equation}
where the sum is over nearest neighbors in the lattice and $g$ is the interaction
strength. Time-evolution of the qubits under this Hamiltonian, $e^{-iHt/\hbar}$,
generates pair-wise controlled-phase operators of the form
\begin{equation} \label{eqn:cs}
CS_{\phi}=\ket{00}\bra{00}+\ket{01}\bra{01}+\ket{10}\bra{10}+e^{i\phi}\ket{11}\bra{11}\mathrm{,}
\end{equation}
with $\phi=gt/\hbar$ proportional to the intensity and duration of the interactions.
Systems in which such techniques have been proposed include quantum
dots~\cite{Borhani2005,Weinstein2005}, superconducting qubits~\cite{Tanamoto2006,You2007}, and
optical lattices~\cite{Duan2003,Garc'ia-Ripoll2003}. An alternate approach has been proposed for
the optical lattice implementation, where by varying the polarizations of the
lattice lasers, the atoms' positions can be shifted state-dependently so as to
induce collisional interactions between neighboring atoms~\cite{Jaksch1999,Mandel2003}.
This results in pair-wise controlled-phase operators of the form
\begin{eqnarray} \label{eq:CSX}
CSX_{\phi}&=&(I\otimes\sigma_x)CS_{\phi}(I\otimes\sigma_x) \label{eqn:csx} \\
&=&\ket{00}\bra{00}+\ket{01}\bra{01}+e^{i\phi}\ket{10}\bra{10}+\ket{11}\bra{11}\mathrm{,}\nonumber
\end{eqnarray}
with $\phi$ again proportional to the intensity and duration of the
interactions.

Ideally, phases of exactly $\phi=\pi$ are applied, resulting in perfect
cluster states ($CZ\equiv CS_{\pi}$). In practice, however, such precision is
impossible: the actual phases are likely to be of the form $\phi=\pi+\theta$,
with $\theta$ a small but unknown systematic phase error. Generally, imperfect
cluster states would result, in which the entanglement between neighboring
pairs of qubits is non-maximal. These systematic phase errors would lead to
unacceptably large fidelity losses during computation for cluster states of
practical sizes, similar to those resulting from the random phase errors
considered by Tame~{\it et al.}~\cite{Tame2005}. While such random errors may
also be present, they would likely be narrowly distributed about the
systematic error value $\theta$. Although it has been shown that standard fault tolerance schemes can be
applied to 1WQC~\cite{Nielsen2005a,Aliferis2006}, a more direct approach to removing these non-separable correlated errors would allow greater efficiency.

This issue has been previously considered in the context of NMR~\cite{Jones2003,Brown2004}; composite pulse sequences were proposed as a means for reducing the effects of systematic phase errors in two-qubit gates. Originally developed to reduce the impact of rotation-angle errors in single-qubit gates, composite pulse sequences were shown by Jones~\cite{Jones2003} to be analogously applicable to phase errors in two-qubit gates generated from imperfectly controlled Ising interactions. The composite pulse approach was further generalized by Brown~{\it et al.}~\cite{Brown2004} to allow two-qubit gates to be performed with arbitrary accuracy, though this required arbitrarily long composite pulse sequences. Although these investigations were framed in the context of NMR and the conventional circuit model of quantum computing, it can in principle be applied to the generation of cluster states in the various aforementioned physical implementations. Despite this, the work of Tame~{\it et al.}~\cite{Tame2005} suggests a practical technique for complete removal, rather than reduction, of systematic phase errors is desirable.

\section{Stochastic teleportation}
We present a technique for producing perfect cluster states despite the
presence of systematic phase errors in the entangling process, thereby
allowing high-fidelity 1WQC. Our approach is based on the use of a stochastic
protocol for restoring maximal entanglement via measurements, together with
multiple applications of an imperfect global entangling operation, each with a distinct value of $\theta$.
An array of perfect two-qubit cluster states is initially distilled from the
improperly prepared initial state. These two-qubit cluster states are then
fused together using previously developed techniques~\cite{Browne2005,Barrett2005,Benjamin2005} to produce a single cluster state of arbitrary size.
Algorithm-specific graph states can be constructed directly, as can be more
exotic graph states.

\subsection{One-bit teleportation}
In 1WQC, logical qubits are represented by entangled chains of physical
qubits (one-dimensional cluster states). Unitary operations on the logical
qubits are effected through repeated use of the `one-bit teleportation'
primitive~\cite{Zhou2000,Nielsen2005}. The left-most qubit, unlike all other qubits in
the chain, is initially in the arbitrary input state $\ket{\psi}$ instead of
the $\ket{+}$ state, and is measured in the $\xi$-basis with outcome
$\ket{m}$. In practice, this can be achieved by first applying a rotation of
angle $\xi$ about the $\sigma_z$-axis, then applying a Hadamard operator, and
finally performing a measurement in the $\sigma_z$-basis. The $\xi$-basis has
eigenstates $(\ket{0}\pm e^{-i\xi}\ket{1})/\sqrt{2}$, which lie in the
$\sigma_x\sigma_y$ plane of the Bloch sphere, rotated from the $\sigma_x$-axis
by angle $\xi$ about the $\sigma_z$-axis. As a result of this measurement, the state
\begin{equation} \label{eq:onebitout}
  \sigma_x^m HR_z(\xi)\ket{\psi}
\end{equation}
is teleported to the next qubit along the chain, and acts as the input state for the following one-bit teleportation. 1WQC works by exploiting the universality of the operator $HR_z(\xi)$ for single qubit operations.

When one-bit teleportation is performed on an imperfect cluster state, the
output state is no longer $\sigma_x^m HR_z(\xi)\ket{\psi}$, but rather a
$\theta$-dependent state that cannot be expressed as $U\ket{\psi}$ for some
unitary operator $U$ that is independent of the input state. We therefore
refer to this as a non-unitary distortion, in the sense that different input
states are not acted upon by the same unitary operator. The resulting fidelity
loss, averaged over all possible input states and both possible measurement
outcomes, is $\frac{1}{2}\sin^2(\theta/2)$. Although this fidelity loss will
be small provided $\theta$ is expectedly small, any practical quantum
algorithm will necessitate a long sequence of concatenated one-bit
teleportations, over which these fidelity losses will rapidly build
up~\cite{Tame2005}.

\subsection{The three-qubit stochastic protocol}
\begin{figure}[t]
  \begin{center}
  \includegraphics[width=0.6\textwidth]{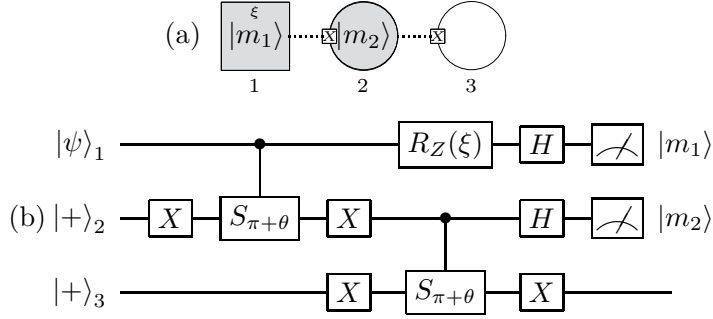}
  \end{center}
\caption{(a) Graph representation of the stochastic protocol.
A square node denotes a qubit in state $\ket{\psi}$, while a circular node
denotes a qubit in state $\ket{+}$. A dotted edge with an ``$X$'' at the
target qubit denotes a $CSX_{\pi+\theta}$ operator. Shaded qubits are measured
in the $\sigma_x$-basis, unless an arbitrary basis angle ($\xi$) is indicated.
The measurement outcome is indicated by the contents of a node. (b) The
equivalent quantum circuit.}
\label{fig:three}
\end{figure}

Consider instead an imperfect three-qubit chain built from $CSX_{\pi+\theta}$ operators (\ref{eq:CSX}) instead of $CZ$ operators, where qubit~1 is in the arbitrary state $\ket{\psi}=\alpha\ket{0}+\beta\ket{1}$ and
the remaining qubits are in $\ket{+}$. In its simplest form, the stochastic
protocol consists of measuring the middle qubit in the $\sigma_x$-basis, as
shown in Fig.~\ref{fig:three}. If the measurement outcome is $\ket{+}$ ($m_2=0$), then
the protocol fails, leaving the two unmeasured qubits in the non-maximally
entangled state 
\begin{equation} \label{eq:fail}
\frac{\left(1-e^{i\theta}\right)\alpha}{4}\ket{00}+\frac{\alpha}{2}\ket{01}
-\frac{e^{i\theta}\beta}{2}\ket{10}+\frac{\left(1-e^{i\theta}\right)\beta}{4}
\ket{11}\mathrm{,} \nonumber
\end{equation}
up to a normalization factor. However, if the measurement outcome is
$\ket{-}$ ($m_2=1$), then the protocol succeeds, leaving the two unmeasured qubits in
the state
\begin{equation} \label{eq:succ}
e^{i\theta/2}(\alpha\ket{00}-\beta\ket{11})
=e^{i\theta/2}(I\otimes\sigma_zH)CZ\ket{\psi}\ket{+},
\end{equation}
where the phase error has been transformed into a harmless global phase.
Success is flagged by the measurement outcome and occurs with probability
$\frac{1}{2}\cos^2(\theta/2)$, which remains finite no matter how weak the
entanglement is, provided $\theta\neq\pi$.

The stochastic protocol allows perfect quantum teleportation to be performed
via the one-bit teleportation primitive that is central to 1WQC. If $m_2=1$,
a subsequent measurement of qubit~1 in the $\xi$-basis with outcome
$\ket{m_1}$ [Fig.~\ref{fig:three}] results in the state
\begin{equation} \label{eq:stochout}
  \sigma_z^{(1-m_1)}R_z(\xi)\ket{\psi}
\end{equation}
being teleported to qubit~3 with perfect fidelity. This use of the stochastic
protocol shares many common features with the probabilistic quantum
teleportation proposed by Agrawal and Pati~\cite{Agrawal2002} as a
modification of the original quantum teleportation protocol of 
Bennett~\textit{et al.}~\cite{Bennett1993}. Their scheme also succeeds with a
certain probability and flags success via the measurement outcome; however,
it requires the degree of non-maximal entanglement to be known in advance, so
that a two-qubit measurement in the commensurate non-maximally entangled basis
can be performed. With our stochastic protocol, this two-qubit
measurement is replaced by a $CSX_{\pi+\theta}$ operator with unknown
$\theta$, followed by two single-qubit measurements, one of which ($m_2$) can
be performed in advance to determine whether or not the teleportation will
succeed. It is important to note that the stochastic protocol requires the use
of $CSX_{\phi}$ (not $CS_{\phi}$) operators. These arise naturally from
collisional interactions in optical lattices, but can also be produced from an
Ising interactions with a different configuration of external fields, such
that the minus sign in the second factor of the Hamiltonian~(\ref{eq:onebitout})
is replaced by a plus sign. Furthermore, if all qubits are initialized in the $\ket{+}$ state, then converting between $CS_{\phi}$ and $CSX_{\phi}$ operators simply requires applying a $\sigma_x$ operator to all qubits (since $\sigma_x$ is the eigenoperator for $\ket{+}$).

\subsection{The $n$-qubit stochastic protocol}
The three-qubit stochastic protocol discussed above can be generalized to an
imperfect chain of any odd number of qubits, where $n$ qubits (all but the
first and last qubits) are measured in the $\sigma_x$-basis. For example, success occurs in a five-qubit chain ($n=3$) with probability
$\frac{3}{8}\cos^4(\theta/2)$, corresponding to measurement outcome sequences
of $\ket{101}$, $\ket{111}$, and $\ket{010}$. In a seven-qubit chain ($n=5$),
success occurs for outcome sequences $\ket{00100}$, $\ket{01001}$,
$\ket{01011}$, $\ket{01110}$, $\ket{10010}$, $\ket{10101}$, $\ket{10111}$,
$\ket{11010}$, $\ket{11101}$, and $\ket{11111}$. The set of successful
sequences for any odd-qubit chain can be constructed using the following two
rules:

(i) Any successful sequence with odd Hamming weight (\emph{i.e.} containing an odd number of 1's) can be sandwiched
between a pair of 0's to yield a larger successful sequence (\emph{e.g.}
$\ket{010}\rightarrow\ket{00100}$)

(ii) Arbitrary single-qubit outcomes can be sandwiched between 1's and/or any successful
sequences that were formed by the previous rule, to yield a larger successful sequence (\emph{e.g.}
$\ket{010x1}\rightarrow\ket{01001}\mathrm{,}\ket{01011}$). Half of all
sequences built from two $n\geq3$ sequences are redundant, while all sequences
built from more than two $n\geq3$ sequences are redundant.

Accounting for redundant sequences, it can be shown from these rules that a successful measurement outcome for (odd) $n$ qubits occurs with probability
\begin{equation} \label{eq:prob}
P_n=\frac{1}{2^{n}}\left(\begin{array}{c} n \\ \frac{n+1}{2} \end{array}\right) \cos^{(n+1)}\left(\frac{\theta}{2}\right) \mathrm{,}
\end{equation}
yielding the two-qubit state
\begin{equation} \label{eq:nout}
e^{i(n+1)\theta/4}(I\otimes\sigma_z^qH)CZ\ket{\psi}\ket{+} \mathrm{,}
\end{equation}
where $q$ is the Hamming weight of the outcome sequence. Using Stirling's
approximation $\ln(n!)\approx n\ln(n)-n+\frac{1}{2}\ln({2\pi n})$ for large
$n$, one finds that $P_n\sim\sqrt{2/\pi n}\cos^{(n+1)}\left(\frac{\theta}{2}\right)$.

Rule (i) can be understood by considering an imperfect $(n+4)$-qubit chain, with two qubits on the left and right separated by odd $n$ qubits. Performing a successful stochastic protocol on these $n$ qubits with outcomes having an odd Hamming weight, and making use of (\ref{eq:nout}), yields the four-qubit state:
\begin{equation}
\!\!\!\!\!\!\!\!\!\!\!\!\!\!\!\!\!\!\!\!\!\!\!e^{i(n+1)\theta/4}(CSX_{\pi+\theta})_{n+3,n+4}(\sigma_zH)_{n+3}CZ_{2,n+3}(CSX_{\pi+\theta})_{1,2}\ket{\psi+++}_{1,2,n+3,n+4}\mathrm{.}
\end{equation}
Measurement of qubits 2 and ($n+3$) in the $\sigma_x$-basis with outcomes $m_2=m_{n+3}=0$ then results in qubits 1 and ($n+4$) sharing the state $e^{i(n+3)\theta/4}(I\otimes\sigma_zH)CZ\ket{\psi}\ket{+}$, consistent with (\ref{eq:nout}).

Rule (ii) can be understood by considering an imperfect chain of $n+n'+3$ qubits, with one qubit at either end, and the two chains of length $n$ and $n'$ separated by one qubit. Performing successful stochastic protocols on the chains of odd length $n$ and $n'$ with outcomes having Hamming weight $q$ and $q'$, respectively, and again making use of (\ref{eq:nout}), yields the three-qubit state:
\begin{equation}
\!\!\!\!\!\!\!\!\!\!\!\!\!\!\!\!\!\!\!\!\!\!\!e^{i(n+n'+2)\theta/4}(\sigma_z^{q'}H)_{n+n'+3}CZ_{n+2,n+n'+3}(\sigma_z^qH)_{n+2}CZ_{1,n+2}\ket{\psi++}_{1,n+2,n+n'+3}\mathrm{.}
\end{equation}
Measurement of qubit ($n+2$) in the $\sigma_x$-basis with arbitrary outcome $m$ then results in qubits 1 and ($n+n'+3$) sharing the state $e^{i(n+n'+2)\theta/4}(I\otimes\sigma_z^{q+q'+m}H)CZ\ket{\psi}\ket{+}$, consistent with (\ref{eq:nout}).

Although it is clearly advantageous to use the shortest possible protocols (\emph{i.e.} small $n$) in the interest of maximizing the success probability, there are circumstances in which longer protocols are required. The most important such circumstance is explained in section~\ref{sec:select}, where odd $n\geq3$ protocols are required for generating selective entanglement from global interactions. Longer protocols are also useful in generating direct entanglement links between spatially distant qubits, and can be used to produce other more exotic graph states, as explained in section~\ref{sec:2d}.

\subsection{Trapped Hadamard operators} \label{sec:trapped}
\begin{figure}[t]
  \begin{center}
  \includegraphics[width=0.6\columnwidth]{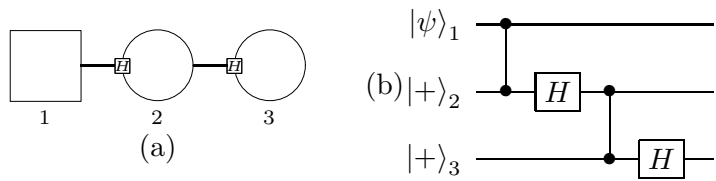}
  \end{center}
\caption{(a) Graph representation of the state resulting from a concatenated sequence of two successful stochastic protocols. A solid edge with an ``$H$'' at the target qubit denotes an $(I\otimes H)CZ$ operator. (b) The equivalent quantum circuit.}
\label{fig:GHZ}
\end{figure}

Although the stochastic protocol allows perfect teleportation,
it is not sufficient for performing universal 1WQC because the output
(\ref{eq:stochout}) lacks the Hadamard operator that is present in the output
of standard one-bit teleportation (\ref{eq:onebitout}). This is due to the
presence of an extra Hadamard in the input state (\ref{eq:succ}), which
cancels the Hadamard in the output. One might na\"\i vely expect that this
extra Hadamard could be eliminated by manually applying a Hadamard to qubit~2
after the stochastic protocol is successfully performed. That this is not
possible is clear when one considers a concatenated sequence of two successful
stochastic protocols, as shown in Fig.~\ref{fig:GHZ}, after which the
remaining three qubits share the state
$H_3CZ_{2,3}H_2CZ_{1,2}\ket{\psi}_1\ket{+}_2\ket{+}_3$ (ignoring $\sigma_z^q$
operators). The Hadamard operating on qubit~2 is trapped between successive
$CZ$ operators, past which a manually applied Hadamard cannot freely commute.

The inability of the stochastic protocol to yield a state that is a universal
entanglement resource for 1WQC can also be understood by noting that when the
input state of qubit~1 is simply $\ket{+}$ and both protocols succeed, the remaining three qubits
share the perfect GHZ state $\left(\ket{000}+\ket{111}\right)/\sqrt{2}$. More
generally, a concatenated sequence of $N$ successful stochastic protocols will
generate a $(2N-1)$-qubit GHZ state. Although Bell states and three-qubit GHZ states can be transformed into two- and three-qubit cluster states via the application of Hadamards, this does not generalize to larger states. For four or more qubits, GHZ
states cannot be transformed into cluster states via any local operations and
classical communication (LOCC)~\cite{Briegel2001}, nor are they sufficient
for 1WQC~\cite{Vandennest2006}.

\section{Selective entanglement from global interactions} \label{sec:select}
\begin{figure}[t]
  \begin{center}
  \includegraphics[width=0.6\columnwidth]{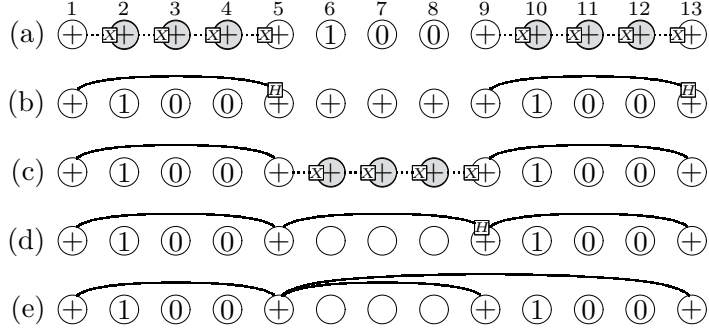}
  \end{center}
\caption{The contents of each node indicate the state of that qubit. (a)~Qubits 6-8 are initialized such that they remain unentangled, leaving two disjoint imperfect five-qubit chains upon which the stochastic protocol is performed until successful. (b)~The resulting Bell pairs are transformed into two-qubit cluster states via the application of Hadamards. Meanwhile, qubits 6-8 are rotated into the $\ket{+}$ state, and qubits 2-4 and 10-12 are rotated such that they will be unaffected by the next entanglement process. (c)~Qubits 5-9 then form an imperfect five-qubit chain, upon which the stochastic protocol is performed until successful. (d)~Qubit 9 (a `tail' qubit) acquires a trapped Hadamard. (e)~By applying a Hadamard to qubit 9 (`tail'), and a $\sigma_z$ operator to qubit 8 (`tip'), a four-qubit `3-node' state results. Qubit 9 (`tail') can be disentangled via measurement in the $\sigma_z$-basis to leave a perfect three-qubit 1D cluster state.}
\label{fig:thirteen}
\end{figure}

Fortunately one can go beyond LOCC by using multiple applications of the
imperfect global entangling process (each with a different value of $\theta$),
together with the stochastic protocol, in such a manner that perfect cluster states 
of arbitrary size are produced by fusing together an array of two-qubit cluster states.
The procedure for a one-dimensional array of 13 qubits is illustrated in
Fig.~\ref{fig:thirteen}, using the $n=3$ protocol, though the procedure can in principle be performed with any odd $n\geq 3$ protocol.

Judiciously choosing the initial states of the
qubits, only select pairs of neighboring qubits are entangled by the
$CSX_{\pi+\theta}$ operators. In particular, neighboring pairs in the states
$\ket{\chi}\ket{1}$ and $\ket{0}\ket{\chi}$ (where $\ket{\chi}$ is an
arbitrary state) will not become entangled, because the phase in the
entangling operator~(\ref{eqn:csx}) acts only on the $\ket{1}\ket{0}$ state.
In this way an array of imperfect five-qubit chains can be created, each
separated by three unentangled qubits [Fig.~\ref{fig:thirteen}(a)].

Successful applications of the stochastic protocol on these chains then yield
multiple unconnected Bell pairs instead of a single large GHZ state
[Fig.~\ref{fig:thirteen}(b)]. Hadamards are then applied manually to the
second qubit of each Bell pair (henceforth referred to as the `tip'
qubit, while the first qubit is designated the `tail'). This transforms each Bell pair into an isolated two-qubit cluster
state (equivalent to removing an untrapped Hadamard).

Meanwhile, the three qubits bewteen neighboring two-qubit cluster states are rotated into $\ket{+}$. 
Again applying the selective entangling technique [Fig.~\ref{fig:thirteen}(c)], the previously measured qubits neighboring the tip and tail qubits are rotated into states $\ket{0}$ and $\ket{1}$, respectively. Subsequent application of the global entangling operation then connects the neighboring two-qubit cluster states `tip-to-tail' by imperfect five-qubit chains, without generating any additional unwanted entanglement to the previously measured qubits. Successful applications of the stochastic protocol
on these imperfect connecting chains then serve to fuse the two-qubit cluster states
together [Fig.~\ref{fig:thirteen}(d)]. The illustrated four-qubit state produced by fusing a pair of two-qubit cluster states is
\begin{equation}
 \ket{\Psi}_{1,5,9,13}=CZ_{9,13}H_9CZ_{5,9}CZ_{1,5}\ket{++++}_{1,5,9,13}\mathrm{.}
\end{equation}

This last step does not quite result in the larger cluster state being sought.
Rather, the left (tail) qubit of each two-qubit cluster state acquires a trapped Hadamard upon being
successfully fused to the right (tip) qubit of an adjacent two-qubit cluster state. 
This state can be transformed into a graph state by subsequently applying Hadamards to these tail qubits, and $\sigma_z$ operators to the commensurate tip qubits. The resulting graph consists of a 1D cluster state composed of former tip qubits, each of which is also connected to a former tail qubit of vertex degree one [qubit 9 in Fig.~\ref{fig:thirteen}(e)]. These former tail qubits (which are designated `leaf' qubits in
the discussion below) can then be disentangled and thereby removed from the perfect 1D cluster state by measuring them in the $\sigma_z$ basis.

\subsection{Fail and retry} \label{sec:failretry}
An important feature of the selective entangling technique described above
is that the stochastic protocol can be retried when it fails.
If the protocol fails on a given $(n+2)$-qubit chain, the middle $n$ qubits can
be re-initialized into $\ket{+}$ states, subsequently re-entangled by the
global $CSX_{\pi+\theta}$ operators, and re-measured until a successful
outcome is obtained. Once a two-qubit cluster state has been successfully
created, appropriate re-initialization of the neighboring qubits ensures that
it is not disturbed by subsequent entangling operations used to retry
failed protocols elsewhere in the chain. Likewise, if the protocol fails while attempting to fuse two cluster states (\emph{e.g.} Fig.~\ref{fig:thirteen}), it can be retried.

Remarkably, the above is true no matter what the state of the first and last qubits prior to repeating the stochastic protocol (\emph{e.g.} states resulting from failed protocols). If the first and last qubits are initially in the arbitrary two-qubit state $\ket{\Psi}=\alpha\ket{00}+\beta\ket{01}+\gamma\ket{10}+\delta\ket{11}$, with the $n$ qubits between them initialized in the $\ket{+}$ state, and the entire ($n+2$)-qubit chain subsequently entangled by global $CSX_{\pi+\theta}$ operators, then a successful sequence of $\sigma_x$-basis measurement outcomes on the middle $n$ qubits will result in the state $e^{i(n+1)\theta/4}(\alpha\ket{00}+(-1)^q\delta\ket{11})/\sqrt{|\alpha|^2+|\delta|^2}$
being shared by the first and last qubits, where $q$ is the Hamming weight of the outcome sequence. Because this works regardless of the initial state of the first and last qubits, and because the ratio $\alpha/\delta$ is never altered by a failed protocol (\ref{eq:fail}), it does not matter how many times the stochastic protocol fails prior to succeeding. However, the probability of success after $N$ consecutive failures, $P_n^N$, is proportional to $\cos(\theta/2)\sin^{2N}(\theta/2)$ (assuming the same $\theta$ in each global entangling operation), and thus effectively falls to zero after only one failure when $\theta$ is small. For larger values of $\theta$, $P_n^N$ decreases more gradually, but is smaller to begin with. Numerical calculations suggest that for all values of $\theta$, the sum of probabilities converges to the expression for the success probability of a single stochastic protocol (\ref{eq:prob}) in the $\theta=0$ limit,
\begin{equation}
 \sum^{\infty}_{N=0}{P_n^N}=\frac{1}{2^{n}}\left(\begin{array}{c} n \\ \frac{n+1}{2} \end{array}\right)\mathrm{,}
\end{equation}
falling far short of 1, and attaining a maximum of $1/2$ for $n=1$.

\begin{figure}[t]
  \begin{center}
  \includegraphics[width=0.7\columnwidth]{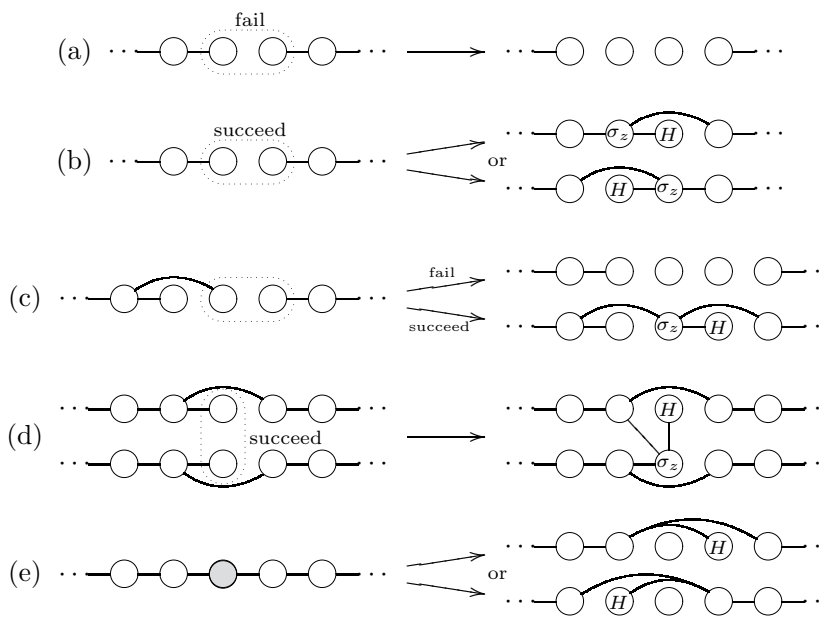}
  \end{center}
\caption{Use of the stochastic protocol as a fusion operation. (a) When the protocol fails, the two directly involved qubits are measured in the $\sigma_z$ basis, leaving perfect cluster states that are each one qubit smaller. (b) When the protocol succeeds, a Hadamard is applied to one of the two directly involved qubits, which thereby becomes a `leaf' qubit. A $\sigma_z$ operator is applied to the other qubit. (c) A failed fusion operation does not reduce the `length' of a growing cluster state (left) that ends with two consecutive leaf qubits. (d) Successful fusion of leaf qubits from adjacent horizontal chains results in a vertical link. (e) $\sigma_x$-basis measurements can be used to create leaf qubits and reduce the length of horizontal chains.}
\label{fig:fusion}
\end{figure}

This inevitability of failure is not a problem when creating an isolated two-qubit cluster state, since all of the qubits involved can simply be measured, re-initialized, and re-entangled. Unfortunately, the inevitability of failure presents a serious problem when attempting to fuse two cluster states together to produce a larger cluster state, since the qubits comprising the cluster states cannot be measured without destroying their entanglement. However, by performing $\sigma_z$-basis measurements on only the two qubits directly involved in a failed fusion attempt (tip and tail), they become disentangled from their respective cluster states, each of which despite being one qubit smaller, remain perfect cluster states. This fusion operation, illustrated in Fig.~\ref{fig:fusion}(a) and \ref{fig:fusion}(b), is nearly identical to the entanglement operation proposed by Barrett and Kok~\cite{Barrett2005}, and the fusion gates proposed by Browne and Rudolph~\cite{Browne2005}.

\subsection{Growth of 1D cluster states}
Despite the fact that a failed attempt to fuse two cluster states reduces the entanglement of each by one qubit [Fig.~\ref{fig:fusion}(a)], cluster states and other graph states can still be grown efficiently~\cite{Barrett2005,Browne2005,Benjamin2005}. In our appoach, we adopt the growth strategy proposed by Benjamin~\cite{Benjamin2005}, which exploits the so-called `leaf' qubits (referred to as `redundant' qubits in ~\cite{Browne2005}) that are produced from successful fusion operations. A leaf qubit is any qubit in a graph state having a vertex degree of one (\emph{e.g.} qubit 9 in Fig.~\ref{fig:thirteen}(e)). When a fusion operation is successful [Fig.~\ref{fig:fusion}(b)], a Hadamard can be manually applied to one of the two directly involved qubits (tip or tail), thereby selecting it as a leaf qubit (in Fig.~\ref{fig:thirteen}(e), the tail qubit was selected). At the same time, a $\sigma_z$ operator can be manually applied to the non-leaf qubit so as to remove a $\sigma_z$ byproduct generated by the successful fusion operation. It is interesting to note that the trapped Hadamard of Fig.~\ref{fig:thirteen}(d), which was at first sight a hindrance, is in fact responsible for allowing a leaf qubit to be created from a successful fusion operation.

The growth of a 1D cluster state is achieved by successively attempting to fuse small cluster states of length $\ell$ to the end of the growing cluster state. We define the length $\ell_{\mathcal{C}}$ of the growing cluster state as the number of qubits in the longest linear segment. For example, the length of the 1D cluster state illustrated in Fig.~\ref{fig:thirteen}(e) is 3; not 4. When fusion is successful, the length of the growing cluster state increases by $\ell-1$. However, when fusion fails, the length of the growing cluster state does not necessarily decrease by one. If the last two qubits in the growing cluster state are both leaf qubits, then a failed fusion operation does not result in a decrease in length [Fig.~\ref{fig:fusion}(c)]. By always constructing an appropriate graph structure for the small cluster states (\emph{e.g.} the `3-node' of Fig.~\ref{fig:thirteen}(e)~\cite{Benjamin2005}), successful fusion always results in the growing cluster state having two leaf qubits at its end. In this way, the length of the growing cluster state decreases by one only when two consecutive fusion operations fail.

To ensure consistent net growth of the cluster state, each fusion operation must on average result in an increase in the number of entanglement links. When fusing small cluster states consisting of $l$ links, successful fusion results in an increase of $l+1$ links in the growing cluster state, while failed fusion always results in a loss of one link. The condition for net growth is then found to be
\begin{equation}
  l>\frac{1}{P_n}-2 \mathrm{,}
\end{equation}
so that the minimum number of links required in the small cluster states depends on the success probability (\ref{eq:prob}), and hence, on the magnitude of $\theta$. It is not known, however, what size and shape of small cluster states provides optimal efficiency for a given value of $P_n$.

\subsection{Growth of 2D cluster states and other graph states} \label{sec:2d}
The first step toward growing 2D cluster states is to grow 1D cluster states along horizontal rows of qubits, vertically separated by the number of qubits $n$ being measured in the stochastic protocol. The fusion operation can then be performed between leaf qubits in adjacent 1D cluster states. Success results in a vertical entanglement link between the 1D cluster states; failure results in both leaf qubits being disentangled from their respective cluster states. As an added bonus, success results in one of the leaf qubits remaining a leaf, which can then be fused to another adjacent 1D cluster state [Fig.~\ref{fig:fusion}(d)]. Interestingly, whereas the tail qubit in a fusion operation must be leaf qubit, the tip qubit can be of arbitrary vertex degree. However, because failure results in complete disentanglement of both the tip and tail qubits, entanglement loss is minimized by only attempting to fuse leaf qubits.

Because this procedure for creating vertical links between horizontal chains requires the leaf qubits in adjacent 1D cluster states to be vertically aligned, it is necessary to judiciously choose the positions of existing leaf qubits, and also to create new ones. During the creation of the 1D cluster states, the choice of leaf qubit (tip or tail) following each successful fusion operation must be made with the intent of maximizing the vertical alignment of leaf qubits. It is also possible to toggle between the two states illustrated in Fig.~\ref{fig:fusion}(b) by applying Hadamard operators to both the tip and tail qubits. Furthermore, as explained in \cite{Browne2005}, leaf qubits can be created anywhere along an existing 1D cluster state by performing $\sigma_x$-basis measurents, which also shorten the length of the cluster state by two qubits [Fig.~\ref{fig:fusion}(e)]. Once a sufficient number of vertical links has been generated between all adjacent horizontal chains, the horizontal lengths between successive vertical links can be shortened via additional $\sigma_x$-basis measurents, and unwanted leaf qubits can be disentangled via $\sigma_z$-basis measurements, so as to obtain a perfect 2D cluster state with qubits of vertex degree 4. Assuming an average horizontal length of three qubits between adjacent vertical links, each qubit in the final 2D cluster state requires an average overhead of $4(n+1)^2$ physical qubits (an $n+1$ by $4(n+1)$ block).

While the above fusion scheme allows for the creation of perfect 2D cluster states, it is far more efficient for the purpose of 1WQC to generate algorithm-specific graph states (also known as `minimal graph states'~\cite{Benjamin2005a}) directly, instead of using $\sigma_z$-basis measurements to carve them from complete 2D cluster states. Selective fusing can also be used to construct other more exotic graph states, particularly those with vertex degrees greater than the lattice coordination number. This can be accomplished by using concatenated stochastic protocols to generate arbitrary $N$-qubit GHZ states (section~\ref{sec:trapped}), which can be converted via local unitaries to so-called `star' graphs (\emph{i.e.} one qubit has vertex degree $N-1$ and all others have vertex degree one), and subsequently fused together. Alternatively, direct entanglement links can be generated to both near and distant qubits by using stochastic protocols of differing $n$.

\subsection{Number of time steps}
Although the optimal strategy for growing cluster states is unknown, we consider here the number of time steps required to construct a perfect $N\times N$ 2D cluster state using Benjamin's $\mathbb{S}$2 strategy~\cite{Benjamin2005}. The first stage involves the growth of adjacent 1D cluster states by fusing together so-called `3-node' cluster states, which are formed when a pair of two-qubit cluster states are successfully fused together [Fig.~\ref{fig:thirteen}]. The second stage involves the formation of vertical links between adjacent 1D cluster states.

Exploiting the inherent parallelism of the global entangling process, a neighboring pair of unconnected two-qubit cluster states can in principle be prepared near the end of a growing 1D cluster state via simultaneous applications of the $n$-qubit stochastic protocol to neighboring $(n+2)$-qubit chains [Fig.~\ref{fig:thirteen}(a),(b) with $n=3$]. Because it is known from measurement outcomes which protocols succeeded, subsequent reattempts of the protocol are performed on only those chains for which all previous protocols have failed. With $p\equiv P_n$ being the probability of success for the $n$-qubit protocol~(\ref{eq:prob}), the probability that preparation of a single two-qubit cluster state will fail $u$ times before succeeding ($u+1$ attempts) is $p(1-p)^u$. Therefore, the average number of simultaneous attempts required to successfully create a neigboring pair of two-qubit cluster states is
\begin{eqnarray}
\overline{s_a}&=&\sum^{\infty}_{u=0}p(1-p)^u\sum^{\infty}_{v=0}p(1-p)^v(\mathrm{max}[u,v]+1) \\ \nonumber
&=&2p^2\sum^{\infty}_{u=0}\sum^{\infty}_{v=u}(1-p)^{u+v}(v+1) - p^2\sum^{\infty}_{u=0}(1-p)^{2u}(u+1) \\
&=&\frac{1}{p}\Big[1+\Big(\frac{1-p}{2-p}\Big)\Big] \mathrm{,} \nonumber
\end{eqnarray}
where indices $u$ and $v$ enumerate the number of attempts on the first and second chains, respectivly. In comparison, an average of $1/p$ attempts would be required to prepare a single two-qubit cluster state, while $2/p$ attempts would be required two produce a neighboring pair of two-qubit cluster states if the protocols could not be applied simultaneously.

The stochastic protocol is then used to attempt to fuse the pair of two-qubit cluster states into a 3-node [Fig.~\ref{fig:thirteen}(c)-(e) with $n=3$]. If the fusion fails, then the pair of neighboring two-qubit cluster states must be re-prepared with an additional $\overline{s_a}$ attempts. On average, $1/p$ attempts of this fusion operation are required for the formation of a 3-node. The average number of applications of the stochastic protocol required to produce a 3-node is therefore
\begin{eqnarray}
\overline{s_b}&=&\frac{1}{p}\overline{s_a} \\
&=&\frac{1}{p^2}\Big[1+\Big(\frac{1-p}{2-p}\Big)\Big] \mathrm{.} \nonumber
\end{eqnarray}

Once a 3-node has been produced, one last stochastic protocol is used to attempt to fuse it to the growing 1D cluster state, for an average total of $\overline{s_b}+1$ stochastic protocols per attempted fusion. If the fusion succeeds, the growing cluster state increases in length by 2 qubits; if the fusion fails, the end qubit of the growing cluster state is disentangled, though this only results in a loss of length if two consecutive fusion attempts fail [Fig.~\ref{fig:fusion}(c)]. The increase in length per fusion attempt must therefore be averaged over the four possible outcomes from two consecutive fusion attempts (zero, one, or two failures), giving
\begin{eqnarray}
2\overline{\Delta\ell_{\mathcal{C}}}&=&p^2[2(\ell-1)]+2p(1-p)[\ell-1]+(1-p)^2[-1] \\
\overline{\Delta\ell_{\mathcal{C}}}&=&p\ell-\frac{1}{2}(1+p^2) \mathrm{,} \nonumber
\end{eqnarray}
where $\ell=3$ for the 3-node. The total average number of applications of the stochastic protocol required to produce a 1D cluster state of length $\ell_{\mathcal{C}}$ can then be calculated as
\begin{equation}
\overline{s_{1D}}=\frac{\ell_{\mathcal{C}}}{\overline{\Delta\ell_{\mathcal{C}}}}(\overline{s_b}+1) \mathrm{.}
\end{equation}

Each simultaneous application of the protocol corresponds to a quantum circuit of depth five (\emph{i.e.} consists of four sequential time steps): 1) initialize chain qubits in $\ket{+}$ and surrounding qubits in $\ket{0}$ or $\ket{1}$ for selective entanglement, 2) generate imperfect entanglement via global interactions, 3) apply Hadamard operators to the middle $n$ qubits, 4) measure the middle $n$ qubits in the $\sigma_x$-basis, and 5) apply appropriate unitary operations if successful, otherwise disentangle via $\sigma_z$-basis measurements. The total average number of time steps required to produce a 1D cluster state of length $\ell_{\mathcal{C}}$ is therefore
\begin{equation}
\overline{t_{1D}}=5\overline{s_{1D}} \mathrm{,}
\end{equation}
which for $\ell=3$ (3-node), $n=3$, and a systematic phase error of $\theta=0.3$ ($\approx 10\%$), gives a requirement of $\overline{t_{1D}}\approx 23\ell_{\mathcal{C}}$ time steps.

Again exploiting the inherent parallelism of the global entangling process, an entire array of adjacent 1D cluster states can be generated simultaneously with the same linear overhead calculated above. Stochastic protocols can then be applied simultaneously between pairs of vertically aligned leaf qubits from adjacent 1D cluster states to produce all necessary vertical links in a single step. During the construction of the 1D cluster states, each successful fusion of a 3-node resulted in the creation of two leaf qubits in addition to the two-qubit length increase. However, prior to every such fusion, the growing cluster state was equally likely to end in one leaf qubits or two leaf qubits. We therefore estimate that $\ell_{\mathcal{C}}/2$ leaf qubits are connected to each 1D cluster state of length $\ell_{\mathcal{C}}$. To successfully produce $N$ vertical links between a given pair of 1D cluster states, an average of $N/p$ attempts of the stochastic protocol, and hence pairs of vertically aligned leaf qubits, are required. Ignoring the use of $\sigma_x$ measurements to create additional leaf qubits [Fig.~\ref{fig:fusion}(e)], each 1D cluster state must therefore have a length of at least $\ell_{\mathcal{C}}=2N/p$.

This last step of simultaneously creating all necessary vertical links via stochastic protocols again corresponds to a quantum circuit of depth five. Afterward, all horizontal lengths between successive vertical links can be shortened via $\sigma_x$-basis measurements and subsequent application of local unitaries, corresponding to a quantum circuit of depth three: 1) apply Hadamard operators, 2) measure in the $\sigma_x$-basis, and 3) apply appropriate unitary operations to neighboring qubits. The final step, corresponding to a quantum circuit of depth two, simply consists of disentangling all remaining leaf qubits by measuring them in the $\sigma_z$-basis and applying appropriate unitary operations to neighboring qubits. Thus, the transformation from an array of sufficiently long 1D cluster states to a 2D cluster state can in principle be achieved in only 10 time steps, independent of the size $N$.

The final average number of time steps required to generate a perfect 2D $N\times N$ cluster state is therefore
\begin{eqnarray}
\overline{t_{2D}}&=&\overline{t_{1D}}+10 \\
&=&\Big[\frac{10}{p\overline{\Delta\ell_{\mathcal{C}}}}(\overline{s_b}+1)\Big]N+10 \mathrm{,}
\end{eqnarray}
which for $\ell=3$ (3-node), $n=3$, and a systematic phase error of $\theta=0.3$ ($\approx 10\%$), gives a requirement of $\overline{t_{2D}}\approx 65N+10$ time steps. Though the final 2D cluster state consists of $N^2$ qubits, the overhead in time steps only scales as $\sim N$ due to the inherent parallelism of the global entangling process. This provides an advantage over similar optical strategies for which the overhead in time steps scales as $\sim N^2$~\cite{Barrett2005,Benjamin2005,Browne2005}, although perfectly controlled global interactions (\emph{i.e.} $\theta=0$) would in principle allow the construction of an $N\times N$ cluster state in constant time.

\section{Conclusion}

We have presented a stochastic measurement protocol, together with a technique for selectively entangling qubits, that enables the efficient growth of perfect cluster states and other perfect graph states even though the global entangling operator is always imperfect. The method extends previous investigations of probabilistic cluster-state growth in the context of linear optics~\cite{Barrett2005,Browne2005,Benjamin2005}. This approach facilitates high-fidelity 1WQC in a variety of physical implementations, at the cost of a constant overhead in both time steps and physical qubits.

The stochastic protocol differs from the composite pulse approach of~\cite{Jones2003,Brown2004} in that it allows systematic phase errors to be completely eliminated with single-qubit measurements, as opposed to being asymptotically reduced with long composite pulse sequences. However, whereas the stochastic protocol only eliminates systematic phase errors, having the same value between successive pairs of nearest-neighbor qubits, the composite pulse approach can also reduce the impact of random phase errors that vary between successive pairs. In principle, the composite pulse approach could be used in combination with the stochastic protocol to initially reduce the magnitude of both random and systematic phase errors. This would increase the success probability of the stochastic protocol, which could then be used to completely eliminate the remaining systematic phase errors.

\ack
The authors are grateful to Nathan Babcock, Jop Briet, Hilary Carteret, Peter
H{\o}yer, Mehdi Mhalla, Simon Perdrix, Ren{\'e} Stock, and Mark Tame for
stimulating discussions. This work was supported by the Alberta Ingenuity
Fund, the Natural Sciences and Engineering Research Council of Canada, and the
Canada Foundation for Innovation.


\section*{References}

\bibliography{Bibliography}
\bibliographystyle{unsrt}


\end{document}